\documentclass[epj,nopacs]{svjour}
\usepackage{amsfonts,graphics}

\newcommand{\ud}{\mathrm{d}}
\newcommand{\ue}{\mathrm{e}}
\newcommand{\xt}{\vec{x},t}
\newcommand{\xvt}{\vec{x},\vec{v},t}
\newcommand{\xyt}{\vec{x},\vec{y},t}

\begin{document}
\title{Point-source dispersion of quasi-neutrally-buoyant inertial particles}
%\subtitle{}
\author{Marco Martins Afonso \and S\'{\i}lvio M.A. Gama%\inst{1}% etc
%% \thanks is optional - remove next line if not needed
}                     % Do not remove
%\mail{}
%
\institute{Centro de Matem\'atica da Universidade do Porto, Rua do Campo Alegre 687, 4169-007 Porto, Portugal\\\email{marcomartinsafonso@hotmail.it}}
\date{\today}
\abstract{
We analyze the evolution of the distribution, both in the phase space and in the physical space,
of inertial particles released by a spatially-localized (punctual) source and advected by an incompressible flow.
The difference in mass density between fluid and particles is assumed as small, and represents the basic parameter
for a regular perturbative expansion. By means of analytical techniques such as Hermitianization,
we derive a chain of equations of the advection--diffusion--reaction type, easily solvable at least numerically.
Our procedure provides results also for finite particle inertia, away from the over-damped limit of quasi-tracer dynamics.
%
%\PACS{
%      {PACS-key}{text of that key}   \and
%      {PACS-key}{text of that key}
%     } % end of PACS codes
} %end of abstract
\maketitle
\section{Introduction} \label{intro}
Particles transported by a fluid flow are dubbed ``inertial'' if, when investigating their movement,
one cannot disregard the particle relative inertia with respect to the carrier fluid.
This is usually due to a mismatch between the two mass densities, and/or to the (small but) not negligible particle size.
Practical examples are constituted by droplets in gases, small bubbles in liquids, and solid powders in a generic fluid.
Understanding the motion of these inclusions is still an open problem from the numerical, experimental, and theoretical points of view
\cite{BFF01,WM03,B05,CBBBCLMT06,BCH07,FMPV07,VCVLMPT08}.

Under investigation here is the evolution of the particle concentration following a point-source release.
This is especially important to study micro-particles emitted by a syringe in a micro-channel \cite{MPM14},
or particulate matter (PM) inserted from a smokestack into the atmosphere \cite{BBCCMMPP14},
with direct applications in terms of pollutant dispersion, climate change, hydrological cycles,
epidemiology and environmental sciences.
Several results have obtained for tracers, \emph{i.e.} particles behaving as tiny fluid parcels when their inertia is negligible
\cite{CMAM2006,CMAM2007,CMAM07,T18}.
Consequently, many findings exist in literature also for small-inertia particles \cite{MAM11},
which give rise to the so-called ``over-damped'' dynamics
where small deviations take place with respect to the corresponding fluid trajectories.

Our main aim is to propose a new type of expansion which allows one to focus
on regimes arbitrarily far from the over-damped regime to determine the particle concentration.
The expansion parameter is the departure from the limit of neutrally-buoyant particles \cite{MAG17,MAMGM17}.
Such particles are particularly relevant in biophysical applications,
where most of the aquatic microorganisms \cite{LMAFMS12} have a mass density very close to the one of water.
The principal advantage of our expansion
is that the small-scale degrees of freedom can be treated by means of a regular perturbation
theory, rather than by a multi-scale expansion as it happens for the usual over-damped expansion.

In our approach we take into account both particle and fluid inertia (through viscous drag and added mass), gravity and Brownian diffusivity;
we are able to deal with a generic incompressible flow and we find results \emph{a priori}.
Other well-known models from the scientific literature, such as the mesoscopic approach (equations for moments
derived from an integration of the particle density) \cite{SFL02,FSS05} and the kinetic approach (counterpart of the
Maxwell--Boltzmann equation) \cite{R91,R92} seem more suitable for specific applications,
but require some \emph{ad hoc} closure scheme (such as the Chapman--Enskog approximation for the third-order moment),
or can be applied only to simple reference flows, \emph{e.g.} shear/straining ones \cite{R05,SARD09} or steady/plane/parallel ones \cite{S88,S90}.
The technique of Hermitianization has already been used to describe the problem of contaminant dispersion \cite{S85},
but for the case of an instantaneous release, \emph{i.e.} a Dirac delta in time rather than in space.

The paper is organized as follows.
In section \ref{sec}, we sketch the problem under investigation and we show our assumptions.
We subsequently expose our analytical procedure in section \ref{sez}.
Section \ref{sect} is devoted to enounce and discuss the explicit results for specific cases.
Conclusions and perspectives follow in section \ref{concl}.
\section{Statement of the problem} \label{sec}
Let us consider a dilute suspension of tiny identical inertial particles transported by a $d$-dimensional incompressible fluid flow $\vec{u}(\xt)$,
subjected to Brownian diffusion and to a constant gravitational acceleration $\vec{g}$,
and emitted by a spatially-localized source. For instance, an injection point (a syringe) in a microchannel
or a chimney releasing some pollutant in the atmosphere can be approximated as punctual, and placed in the origin of our frame of reference.
If the interactions with physical boundaries or other particles are negligible, the so-called one-way coupling scheme can be implemented,
and the (second-order) dynamics of a single particle can be described by recasting Newton's law into a set of
coupled stochastic differential equations for the particle position $\vec{\mathcal{X}}(t)$ and covelocity $\vec{\mathcal{V}}(t)$:
\begin{equation} \label{basic} 
 \left\{\begin{array}{rcl}
   \dot{\vec{\mathcal{X}}}(t)&=&\vec{\mathcal{V}}(t)+\beta\vec{u}(\vec{\mathcal{X}}(t),t)+\sqrt{2D}\vec{\mu}(t)\;,\\[0.2cm]
   \dot{\vec{\mathcal{V}}}(t)&=&\displaystyle-\frac{\vec{\mathcal{V}}(t)-(1-\beta)\vec{u}(\vec{\mathcal{X}}(t),t)}{\tau}\\
   &&\displaystyle+(1-\beta)\vec{g}+\frac{\sqrt{2\kappa}}{\tau}\vec{\nu}(t)\;.
  \end{array}\right.
\end{equation}
System (\ref{basic}) is based on the results of \cite{G83,MR83},
but here neglected are the corrections due to Fax\'en, Basset, Saffman and Oseen,
while in full generality we take into account the possibility of having Brownian motion act in both evolution equations \cite{Z73,R88,CK98,MAG17,MAMGM17},
through the independent vectorial white noises $\vec{\mu}(t)$ and $\vec{\nu}(t)$ and the corresponding diffusion coefficients $D$ and $\kappa$.
The Stokes time $\tau$ in the denominator of the contribution from viscous drag measures the characteristic delay in the response
of particle dynamics to flow variations, and is given by $\tau\equiv R^2/(3\gamma\beta)$ for spherical particles
of radius $R$ advected by a fluid with kinematic viscosity $\gamma$. Finally, in its turn, the non-dimensional quantity $\beta$
is the added-mass coefficient defined as $\beta\equiv3\rho_{\mathrm{f}}/(\rho_{\mathrm{f}}+2\rho_{\mathrm{p}})$,
starting from the constant mass densities of fluid ($\rho_{\mathrm{f}}$) and particle ($\rho_{\mathrm{p}}$), respectively.
Such a factor ranges from $\beta=0$, for very heavy particles such as droplets or solid powders in a gas, to $\beta=3$,
for very light ones such as bubbles in a liquid; and equals unity, $\beta=1$, for tracers or neutrally-buoyant inclusions.
It causes a macroscopic discrepancy between the particle velocity $\dot{\vec{\mathcal{X}}}(t)$ and covelocity $\vec{\mathcal{V}}(t)$,
and models in an effective fashion the feedback of the particle on the flow, since by definition the velocity field $\vec{u}(\vec{\mathcal{X}}(t),t)$
is the unperturbed one --- that is, the fluid flow computed in the absence of the particle and thus known --- sampled on the particle trajectory.

A statistical average of (\ref{basic}) on the stochastic standard noises \cite{C43,G85,R89,V07}
leads to the generalized Fokker--Planck equation for the phase-space density $p(\xvt)$:
\begin{eqnarray} \label{gfpkfk}
 \left\{\partial_t+\vec{\partial}_{\vec{x}}\cdot[\vec{v}+\beta\vec{u}(\xt)]+\vec{\partial}_{\vec{v}}\cdot\left[\frac{(1-\beta)\vec{u}(\xt)-\vec{v}}{\tau}\right.\right.\nonumber\\
 +(1-\beta)\vec{g}\bigg]-D\partial^2_{\vec{x}}-\frac{\kappa}{\tau^2}\partial^2_{\vec{v}}\bigg\}p(\xvt)=S(\xvt)\;.
\end{eqnarray}
Let us denote with $\mathcal{L}_{\xvt}$ the linear second-order differential operator in curly braces on the left-hand side of (\ref{gfpkfk}),
so that $\mathcal{L}_{\xvt}p=S$. The forcing term on the right-hand side models the particle source, and is therefore proportional to a spatial Dirac delta
(this feature will be discussed in the conclusions in order to take into account also distributed sources) \cite{MAM11,CMAM2006,CMAM2007,CMAM07,T18}.
If one considers an emission with constant rate $T^{-1}$ and a known distribution $f(\vec{v})$ in the covelocity coordinate, then:
\begin{equation} \label{source}
 S(\xvt)=\frac{\delta(\vec{x})f(\vec{v})}{T}\;.
\end{equation}
The normalization condition
\begin{equation} \label{norm}
 \!\int\!\ud\vec{v}\,f(\vec{v})=h
\end{equation}
typically implies $h=1$, but we will show that our formalism succeeds in describing also the case $h=0$, mimicking a forcing term
which acts as a source of particles with some ranges of covelocity and as a sink for some other values of the latter.
Restrictions on the forms of $f(\vec{v})$ tractable analytically in this framework will be imposed later,
for the time being the only necessary condition consisting in a sufficiently-rapid decay at infinity.

Let us define the physical-space concentration as
\begin{equation} \label{phys}
 q(\xt)\equiv\!\int\!\ud\vec{v}\,p(\xvt)\;.
\end{equation}
As an aside remark, a simple integration of (\ref{gfpkfk}) on the whole covelocity space gives:
\begin{eqnarray} \label{iso}
 [\partial_t+\beta\vec{u}(\xt)\cdot\vec{\partial}_{\vec{x}}-D\partial^2_{\vec{x}}]q(\xt)&&\nonumber\\
 +\vec{\partial}_{\vec{x}}\cdot\!\int\!\mathrm{d}\vec{v}\,\vec{v}p(\xvt)&=&\frac{h}{T}\delta(\vec{x})\;.
\end{eqnarray}
This equation is not closed in the physical-space concentration, except when the integral term vanishes.
In particular, in the presence of isotropy for the covelocity variable in both the forcing and the initial/boundary conditions,
only the first line would survive in the left-hand side of (\ref{iso}), and an equation living purely in the physical space would emerge.
\section{Perturbative expansion for quasi-neutral buoyancy and Hermitianization} \label{sez}
The resolution of (\ref{gfpkfk}) is of course a daunting task, also numerically, because the relevant quantities are defined in the full phase space.
If \emph{e.g.}\ $d=3$, a computational approach would face the enormous difficulty of dealing with $3+3+1=7$ independent coordinates,
something very hard to accomplish. The trick then consists in focusing on situations where some further analytical manipulation is possible,
\emph{viz.}\ in the presence of a small parameter denoting a weak departure from a specific reference case, and in terms of which a perturbative expansion
can be performed. When properly done, this often allows for a separation between the covelocity degree of freedom and the space--time ones,
leading to innocent equations (of advection--diffusion--reaction type) purely living in the physical space and thus easily solvable at least numerically.

Let us then focus on particles whose mass density differs mildly (in either shortfall or excess) from the fluid one \cite{BCPP00,MFS07,SH08,MAG17,MAMGM17}.
As $\beta\simeq1$, and $1-\beta$ is small but with an undefined sign, we introduce a positive small parameter in the form of $\alpha\equiv|1-\beta|\ll1$.
We also define $J\equiv\mathrm{sgn}(1-\beta)$, so that $\beta=1-J\alpha$. It can be shown that in this situation it is possible to proceed analytically
only if one makes the further assumption that the Brownian-diffusion coefficient $\kappa$ appearing in the equation for the particle acceleration be small
as well, namely with the same asymptotic behavior as the mass-density mismatch: $\kappa\sim\alpha\ll1$; or, in other words, one can define a finite
constant $K\equiv\kappa/|1-\beta|=\alpha^{-1}\kappa$ with dimensions of square length over time. Notice that no assumption is needed to be made on the
Brownian diffusivity $D$ driving the particle velocity, which can then be thought of as a regularizing parameter, as long as it is non-zero.
 
After rescaling the covelocity variable in the form $\vec{v}\mapsto\vec{y}\equiv\vec{v}/\sqrt{|1-\beta|}=\alpha^{-1/2}\vec{v}$, the generalized
Fokker--Planck operator splits into
\begin{equation} \label{l012}
 \mathcal{L}=\mathcal{L}^{(0)}+\alpha^{1/2}\mathcal{L}^{(1)}+\alpha\mathcal{L}^{(2)}\;,
\end{equation}
with:
\begin{eqnarray*}
 \mathcal{L}^{(0)}&=&\partial_t+\vec{u}(\xt)\cdot\vec{\partial}_{\vec{x}}-D\partial^2_{\vec{x}}-\tau^{-1}\vec{\partial}_{\vec{y}}\cdot\vec{y}-K\tau^{-2}\partial^2_{\vec{y}}\;,\\
 \mathcal{L}^{(1)}&=&\vec{y}\cdot\vec{\partial}_{\vec{x}}+J[\tau^{-1}\vec{u}(\xt)+\vec{g}]\cdot\vec{\partial}_{\vec{y}}\;,\\
 \mathcal{L}^{(2)}&=&-J\vec{u}(\xt)\cdot\vec{\partial}_{\vec{x}}\;.
\end{eqnarray*}
For the sake of notational simplicity, we define a ``gravitational velocity field'' $\vec{z}(\xt)\equiv\vec{u}(\xt)+\tau\vec{g}$ and two linear operators,
\begin{equation} \label{m}
 \mathcal{M}_{\xt}\equiv\partial_t+\vec{u}(\xt)\cdot\vec{\partial}_{\vec{x}}-D\partial^2_{\vec{x}}
\end{equation}
(generator of advection--diffusion in physical space) and
\begin{equation} \label{n}
 \mathcal{N}_{\vec{y}}\equiv\vec{\partial}_{\vec{y}}\cdot\vec{y}+K\tau^{-1}\partial^2_{\vec{y}}
\end{equation}
(Ornstein--Uhlenbeck formalism in covelocity variable). In terms of them,
\begin{equation} \label{l01}
 \mathcal{L}^{(0)}=\mathcal{M}_{\xt}-\tau^{-1}\mathcal{N}_{\vec{y}}\;,\quad\mathcal{L}^{(1)}=\vec{y}\cdot\vec{\partial}_{\vec{x}}+J\tau^{-1}\vec{z}(\xt)\cdot\vec{\partial}_{\vec{y}}\;.
\end{equation}
Solenoidality implies $\vec{\partial}_{\vec{x}}\cdot\vec{u}(\xt)=0\Rightarrow\vec{\partial}_{\vec{x}}\cdot\vec{z}(\xt)=0$.
Our choice for rescaling is driven by the close analogy with the situation shown in \cite{MAM11,MACMO2009,MAMM2011,MA08,MAMM12},
where the low-inertia limit was taken.
In that case the small quantity at denominator was the square root of $\tau$, while here it is that of $|1-\beta|$.
The advantage of such a rescaling lies in the fact that it allows for a full decoupling of the rescaled covelocity from the
physical-space dynamics (see the separated form of $\mathcal{L}^{(0)}$ in (\ref{l01})),
and for the resolution of equations based on the operator (\ref{n}) in terms of a basic Gaussian state.
Notice that the same Jacobian factor $\alpha^{d/2}$ appears in both distribution changes $p(\xvt)\mapsto p(\xyt)$ and $f(\vec{v})\mapsto f(\vec{y})$,
so that it can be discarded, and (\ref{gfpkfk}) is still valid in the new rescaled variable.
 
It is now natural to expand the phase-space density and source into two power series in $\sqrt{\alpha}$:
\begin{equation} \label{esp}
 p(\xyt)=\sum_{n=0}^{\infty}\alpha^{n/2}p^{(n)}(\xyt),\ f(\vec{y})=\sum_{n=0}^{\infty}\alpha^{n/2}f^{(n)}(\vec{y}).
\end{equation}
Replacing into (\ref{gfpkfk}) and using (\ref{l012}), we get:
\begin{eqnarray} \label{label}
 \mathcal{L}^{(0)}p^{(0)}&=&\frac{\delta(\vec{x})}{T}f^{(0)}(\vec{y})\;,\nonumber\\
 \mathcal{L}^{(0)}p^{(1)}&=&-\mathcal{L}^{(1)}p^{(0)}+\frac{\delta(\vec{x})}{T}f^{(1)}(\vec{y})\;,\nonumber\\
 \mathcal{L}^{(0)}p^{(n)}&=&-\mathcal{L}^{(1)}p^{(n-1)}-\mathcal{L}^{(2)}p^{(n-2)}+\frac{\delta(\vec{x})}{T}f^{(n)}(\vec{y})\nonumber\\
 &&\qquad(\textrm{for }n\ge2)\;.
\end{eqnarray}
The initial condition, reading $p(\vec{x},\vec{y},0)=p_*(\vec{x},\vec{y})$ in its original form, must as usual be imposed on the zeroth order,
with all the other orders evolving from the nil distribution:
\begin{equation} \label{ini}
 p^{(n)}(\vec{x},\vec{y},0)=p_*(\vec{x},\vec{y})\delta_{n0}\;.
\end{equation}
Also, expression (\ref{norm}) translates into a constraint for every order:
\begin{equation} \label{normal}
 \!\int\!\ud\vec{y}\,f^{(n)}(\vec{y})=h\delta_{n0}\;.
\end{equation}
The physical-space concentration is also automatically expanded as
\begin{equation} \label{exp}
 q(\xt)=\sum_{n=0}^{\infty}\alpha^{n/2}q^{(n)}(\xt),\ q^{(n)}(\xt)=\!\int\!\ud\vec{y}\,p^{(n)}(\xyt),
\end{equation}
with initial condition
\begin{equation} \label{in}
 q(\vec{x},0)=q_*(\vec{x})\equiv\!\int\!\ud\vec{y}\,p_*(\vec{x},\vec{y})\;,\quad q^{(n)}(\vec{x},0)=q_*(\vec{x})\delta_{n0}\;.
\end{equation}

The following step consists in a Hermitianization of the problem, \emph{i.e.} in an expansion of the relevant quantities
in terms of multivariate Hermite polynomials \cite{G49,S79,S99} times a Gaussian measure, as dictated by the form of (\ref{n}).
Let us denote by $G(\vec{y})$ the normalized centered Gaussian distribution, with variance exactly corresponding to $\sigma^2\equiv K\tau^{-1}$:
\begin{equation} \label{gaus}
 G(\vec{y})\equiv(2\pi\sigma^2)^{-d/2}\ue^{-y^2/2\sigma^2}\;.
\end{equation}
Notice that $\mathcal{N}_{\vec{y}}G(\vec{y})=0$, with such a Gaussian being the only ``admissible'' element in the kernel of this operator on
$\mathbb{R}^d$ (in terms of normalization). The $d$-dimensional Hermite polynomials associated with $G(\vec{y})$ are then defined as
\begin{equation} \label{herm}
 H^{\{m\}}_{i_1\ldots i_m}(\vec{y})\equiv\frac{(-1)^m}{G(\vec{y})}\prod_{l=1}^m\frac{\partial}{\partial y_{i_l}}G(\vec{y})\;,
\end{equation}
with the degree-$m$ polynomial being a symmetric rank-$m$ tensor. For instance, $H^{\{0\}}(\vec{y})=1$, $H^{\{1\}}_i(\vec{y})=y_i/\sigma$,
$H^{\{2\}}_{ij}(\vec{y})=y_iy_j/\sigma^2-\delta_{ij}$, $H^{\{3\}}_{ijk}(\vec{y})=y_iy_jy_k/\sigma^3-(\delta_{ij}y_k+\delta_{jk}y_i+\delta_{ki}y_j)/\sigma$, and so on. For our scope, the four basic properties of such polynomials 
can conveniently be expressed as follows, descending from orthonormalization,
\[\!\int\!\ud\vec{y}\,\tens{H}^{\{m\}}(\vec{y})G(\vec{y})=\delta_{m0}\;,\]
eigenfunctionality,
\[\mathcal{N}_{\vec{y}}[\tens{H}^{\{m\}}(\vec{y})G(\vec{y})]=-m\tens{H}^{\{m\}}(\vec{y})G(\vec{y})\;,\]
grading,
\[\frac{\partial[H^{\{m\}}_{i_1\ldots i_m}(\vec{y})G(\vec{y})]}{\partial y_{i_{m+1}}}=-\frac{H^{\{m+1\}}_{i_1\ldots i_mi_{m+1}}(\vec{y})G(\vec{y})}{\sigma}\;,\]
and recursion:
\begin{eqnarray*}
 y_{i_{m+1}}H^{\{m\}}_{i_1\ldots i_m}(\vec{y})&=&\sigma\sum_{l=1}^m\delta_{i_{m+1}i_l}H^{\{m-1\}}_{i_1\ldots i_{l-1}i_{l+1}\ldots i_m}(\vec{y})\\
 &&+\sigma H^{\{m+1\}}_{i_1\ldots i_mi_{m+1}}(\vec{y})\;.
\end{eqnarray*}
The key point is that, by construction, quantities of the form $\tens{H}^{\{m\}}(\vec{y})G(\vec{y})$ are eigenfucntions of the operators
$\mathcal{L}^{(0)}$ and $\mathcal{L}^{(2)}$, while the action of $\mathcal{L}^{(1)}$ is simply an alteration of the resulting degree
from $m$ to a combination of $m-1$ and $m+1$. The hierarchy in (\ref{label}) therefore conserves parity.
 
The second natural expansion is thus (assuming Einstein's convention of implicit sum on repeated subscripts):
\begin{equation} \label{somma}
 p^{(n)}(\xyt)=\sum_{m=0}^{\infty}p^{[n,m]}_{i_1\ldots i_m}(\xt)H^{\{m\}}_{i_1\ldots i_m}(\vec{y})G(\vec{y})\;,
\end{equation}
\begin{equation} \label{som}
 f^{(n)}(\vec{y})=\sum_{m=0}^{\infty}f^{[n,m]}_{i_1\ldots i_m}H^{\{m\}}_{i_1\ldots i_m}(\vec{y})G(\vec{y})\;.
\end{equation}
In order for our expansion to succeed, we now have to specify more precisely the constraint on the rapidity of the decay at infinity of the source term
in the covelocity variable. Starting from the well-known condition about the square-integrability of a function in the Gaussian measure for the
expansion in a pure series of Hermite polynomials, we note that our own expansion consists in a product of the latter with the Gaussian function itself.
We deduce that our constraint is $f(\vec{y})\ue^{+y^2/4\sigma^2}\in L^2(\mathbb{R}^d,\ud\vec{y})$, or alternatively
--- for sources only dependent on the modulus $y=|\vec{y}|$ --- is $f(|\vec{y}|)\ue^{+y^2/4\sigma^2}y^{(d-1)/2}\in L^2(\mathbb{R}^+,\ud y)$.
For example, besides the standard Gaussian $\ue^{-y^2/2\sigma^2}$, also functions behaving at infinity as $\ue^{-(y/\sigma)^{2+\epsilon}}$,
or $\ue^{-y^2(1+\epsilon)/4\sigma^2}$, or $\ue^{-y^2/4\sigma^2-y}$, or $\ue^{-y^2/4\sigma^2}y^{-d/2-\epsilon}$ ($\forall\epsilon>0$) are acceptable.
Moreover, expression (\ref{normal}) turns into a constraint only for the coefficients of rank $m=0$, while the others are free:
\begin{equation} \label{normaliz}
 f^{[n,0]}=h\delta_{n0}\;.
\end{equation}
Notice that both $\tens{f}^{[n,m]}$ in (\ref{som}) and $\tens{p}^{[n,m]}(\xt)$ in (\ref{somma}) are tensors of rank $m$, but the former are just numerical
coefficients, while the latter are space-time functions, whose evolution equations in the physical space are the subject of the upcoming material.

Upon defining $\nabla_i\equiv\partial_{x_i}$ and recalling (\ref{m}), the substitution of (\ref{somma}) and (\ref{som}) into (\ref{label}) gives:
\begin{eqnarray} \label{equ}
 &&\left(\mathcal{M}_{\xt}+\frac{m}{\tau}\right)p^{[n,m]}_{i_1\ldots i_m}(\xt)=\nonumber\\
 &&\quad-\sigma\sum_{l=1}^{m+1}\nabla_jp^{[n-1,m+1]}_{i_1\ldots i_{l-1}ji_l\ldots i_m}(\xt)\nonumber\\
 &&\quad-\sigma\nabla_{i_m}p^{[n-1,m-1]}_{i_1\ldots i_{m-1}}(\xt)+\frac{Jz_{i_m}(\xt)}{\tau\sigma}p^{[n-1,m-1]}_{i_1\ldots i_{m-1}}(\xt)\nonumber\\
 &&\quad+J\vec{u}(\xt)\cdot\vec{\nabla}p^{[n-2,m]}_{i_1\ldots i_m}(\xt)+\frac{\delta(\vec{x})}{T}f^{[n,m]}_{i_1\ldots i_m}
\end{eqnarray}
(with the convention of setting to nought all the tensors where at least one superscript $n$ or $m$ is negative).
Knowing all $\tens{f}^{[n,m]}$'s, first of all we should point out what happens for the unforced sectors,
\emph{i.e.}\ those couples of naturals where $\tens{f}^{[n,m]}=0$ and all the $\tens{p}^{[\bullet,\bullet]}(\xt)$'s
appearing as sources on the right-hand side happen to vanish.
In the presence of incompressible flows, it is well known that, after a possible initial transient which is
not the scope of our investigation, the solution of the unforced advection--diffusion--reaction equation
$(\mathcal{M}_{\xt}+m\tau^{-1})\tens{p}^{[n,m]}=0$ decays to zero (for $m>0$); when $m=0$, the unforced
equation is purely of the advection--diffusion type and the solution tends to a constant, whose only admissible value
for $n>0$ is zero because of normalization. As a consequence, the only sectors to be effectively
solved through (\ref{equ}) are the basic one $n=0=m$, those for which $\tens{f}^{[n,m]}\neq0$,
plus those which have non-zero sources $\tens{p}^{[\bullet,\bullet]}(\xt)$ on the right-hand side.
The system (\ref{equ}) can be solved recursively at least numerically,
since all equations are of the advection--diffusion(--reaction) type and are coupled only in one direction (growing $n$).
In the next section, we will show explicit resolutions for some
paradigmatic cases where only a finite set of $\tens{f}^{[n,m]}$'s is non-zero.
If this is the case, let us denote by $m_n$ the maximum $m$ for which the particle source $f$ is non-zero at that particular $n$.

Note that the double expansion in powers of $\sqrt{\alpha}$ and in the Hermite polynomials,
coupled with the behavior of the operators $\mathcal{L}^{(\bullet)}$,
creates a full decoupling between the quantities with even and odd grades,
where the grade of $\tens{p}^{[n,m]}$ is defined as $n+m$.
In other words, if one arranges the $\tens{p}^{[n,m]}$'s on the equivalent of a semi-infinite chessboard,
with the power exponent $n$ on the abscissae and the Hermite rank $m$ on the ordinates,
the coupling schemes somehow remind of the moves of a bishop in chess,
since either of them conserves its light-squared or dark-squared character by moving only in diagonal.
The dark square on the bottom-left corner is the basic state $p^{[0,0]}$.
From (\ref{equ}), it appears that an even-grade quantity on a dark square is forced only by even-grade ones,
in particular directly only by --- at most, if existing --- three dark squares:
the ones sitting on its bottom-left and top-left (for $n\ge1$), plus the second-neighbor on the left (for $n\ge2$);
and it is forced indirectly by all those other dark squares reachable recursively with this procedure.
Of course the same statement holds for the odd counterpart, represented by light squares.
The limit of maximum $n$ --- say $N$ --- up to which one wants to solve
is simply dictated by the precision needed and by the magnitude of $|1-\beta|$.
Once $N$ is fixed, for every $n<N$ one would like to find the maximum $m$ --- say $M_n$ ---
up to which the solving is necessary in the worst scenario, to find the full relevant $p^{(n)}(\xyt)$. 
It turns out that $M_{n=0}=m_{n=0}$ and, recursively, $M_n=\max\{m_n,M_{n-1}+1\}\ \forall n\ge1$.

However, if one is only interested in the physical-space concentration $q(\xt)$,
one can focus just on the bottom rank $p^{[n,0]}(\xt)=q^{(n)}(\xt)$ due to the Hermite integration property,
and for each $n$ one sees that the relevant quantities (appearing as direct or indirect sources)
are the ones living on the same-colored squares with grade $\le n$,
\emph{i.e.}\ contained in the corresponding right triangle with right angle
in the origin and one vertex in the point under observation.
As a consequence, in this case, $M_{n=N}=0$, $M_{n=N-1}=1$, and in the worst scenario $M_n=N-n$ at most.
\section{Results and discussion} \label{sect}
In this section we are going to present illustrative, explicit results for specific, physically-relevant forcing cases,
where the source term is active only on a limited number of sectors in the two aforementioned expansions.
We remind that, in any case, $f^{[n,0]}=0\ \forall n>0$ because of (\ref{normaliz}),
and that the complete forms of source term and particle concentration
are reconstructed through:
\begin{equation} \label{somm}
 f(\vec{y})=\sum_{n=0}^{\infty}\sum_{m=0}^{\infty}\alpha^{n/2}f^{[n,m]}_{i_1\ldots i_m}H^{\{m\}}_{i_1\ldots i_m}(\vec{y})G(\vec{y})\;,
\end{equation}
\begin{equation} \label{sum}
 p(\xyt)=\sum_{n=0}^{\infty}\sum_{m=0}^{\infty}\alpha^{n/2}p^{[n,m]}_{i_1\ldots i_m}(\xt)H^{\{m\}}_{i_1\ldots i_m}(\vec{y})G(\vec{y})\;.
\end{equation}
\subsection{Case $h=1$} \label{c1}
If $h=1$ in (\ref{norm}), then $f^{[0,0]}=1$. Let us study first what happens when this is the only forced sector, \emph{i.e.}\ $f^{[n,m]}=0\ \forall m>0$,
and the source term perfectly coincides with the Gaussian distribution with standard deviation exactly given by $\sigma$: $f(\vec{y})=G(\vec{y})$.
Apart from transient decays, $p^{[0,m]}(\xyt)=0\ \forall m>0$, so that:
\[p^{(0)}(\xyt)=p^{[0,0]}(\xt)H^{\{0\}}(\vec{y})G(\vec{y})=q^{(0)}(\xt)G(\vec{y})\;,\]
with
\begin{equation} \label{nought}
 [\partial_t+\vec{u}(\xt)\cdot\vec{\nabla}-D\nabla^2]p^{[0,0]}(\xt)=\frac{\delta(\vec{x})}{T}\;.
\end{equation}
We find that, at the lowest order in the expansion, the physical-space concentration
follows the same equation as a passive scalar. This is an interesting result because,
while in the limit of small inertia an inertial particle becomes a tracer by definition,
such an identification in principle does not hold in the limit of neutral buoyancy.
To solve the problem up to $O(\alpha)$, the other relevant equations are:
\begin{equation} \label{11}
 \left(\partial_t+\vec{u}\cdot\vec{\nabla}-D\nabla^2+\frac{1}{\tau}\right)p^{[1,1]}_i=\left(\frac{Jz_i}{\tau\sigma}-\sigma\nabla_i\right)p^{[0,0]}\;,
\end{equation}
\begin{equation} \label{20}
 (\partial_t+\vec{u}\cdot\vec{\nabla}-D\nabla^2)p^{[2,0]}=J\vec{u}\cdot\vec{\nabla}p^{[0,0]}-\sigma\nabla_jp^{[1,1]}_j\;,
\end{equation}
\begin{equation} \label{22}
 \left(\partial_t+\vec{u}\cdot\vec{\nabla}-D\nabla^2+\frac{2}{\tau}\right)p^{[2,2]}_{ij}=\left(\frac{Jz_j}{\tau\sigma}-\sigma\nabla_j\right)p^{[1,1]}_i\;;
\end{equation}
these make up $p^{(1)}=p^{[1,1]}_iy_i\sigma^{-1}G(\vec{y})$ and $p^{(2)}=[p^{[2,0]}+p^{[2,2]}_{ij}H^{\{2\}}_{ij}(\vec{y})]G(\vec{y})$
in the phase space, so that $q^{(1)}=0$ and $q^{(2)}=p^{[2,0]}$ in the physical space. If one wanted to solve the problem up to $O(\alpha^2)$,
one should also write down the equations for $p^{[3,1]}_i$, $p^{[3,3]}_{ijk}$, $p^{[4,0]}$, $p^{[4,2]}_{ij}$ and $p^{[4,4]}_{ijko}$,
in order to reconstruct $p^{(3)}=[p^{[3,1]}_iy_i\sigma^{-1}+p^{[3,3]}_{ijk}H^{\{3\}}_{ijk}(\vec{y})]G(\vec{y})$ and
$p^{(4)}=[p^{[4,0]}+p^{[4,2]}_{ij}H^{\{2\}}_{ij}(\vec{y})+p^{[4,4]}_{ijko}H^{\{4\}}_{ijko}(\vec{y})]G(\vec{y})$ in the phase space,
and $q^{(3)}=0$ and $q^{(4)}=p^{[4,0]}$ in the physical space. Then, $q=q^{(0)}+\alpha q^{(2)}+\alpha^2 q^{(4)}+O(\alpha^3)$.
\subsubsection{Source acting on a higher Hermite sector} \label{c1h}
Let us suppose that, beyond $f^{[0,0]}=1$, now also $f^{[0,2]}_{ij}=\mathcal{F}_{ij}$, with $\mathcal{F}_{ij}$ a known numerical matrix
with some non-zero entries. This represents an excitation with the same parity in terms of grade as above, but with the presence of some anisotropy.
Then, $q^{(0)}$ does not change, but $p^{(0)}=[p^{[0,0]}+p^{[0,2]}_{ij}H^{\{2\}}_{ij}(\vec{y})]G(\vec{y})$, with
\[\left(\partial_t+\vec{u}\cdot\vec{\nabla}-D\nabla^2+\frac{2}{\tau}\right)p^{[0,2]}_{ij}=\frac{\delta(\vec{x})}{T}\mathcal{F}_{ij}\;.\]
To solve the problem up to $O(\alpha)$ in the physical space, \emph{i.e.}\ to find $q^{(2)}$,
the other relevant equations are the previous one, plus the same (\ref{20}) for $p^{[2,0]}$, plus the new version of (\ref{11}):
\begin{eqnarray*}
 \left(\partial_t+\vec{u}\cdot\vec{\nabla}-D\nabla^2+\frac{1}{\tau}\right)p^{[1,1]}_i&=&\left(\frac{Jz_i}{\tau\sigma}-\sigma\nabla_i\right)p^{[0,0]}\\
 &&-\sigma\nabla_j(p^{[0,2]}_{ij}+p^{[0,2]}_{ji})\;.
\end{eqnarray*}
One also finds that $p^{(1)}=[p^{[1,1]}_iy_i\sigma^{-1}+p^{[1,3]}_{ijk}H^{\{3\}}_{ijk}(\vec{y})]G(\vec{y})$
and that $p^{(2)}=[p^{[2,0]}+p^{[2,2]}_{ij}H^{\{2\}}_{ij}(\vec{y})+p^{[2,4]}_{ijko}H^{\{4\}}_{ijko}(\vec{y})]G(\vec{y})$.
As a consequence, the equations for $p^{[1,3]}_{ijk}$, $p^{[2,2]}_{ij}$ and $p^{[2,4]}_{ijko}$ should be attacked too for the full phase-space solution.
\subsubsection{Source acting on a higher perturbative order} \label{c1p}
Let us suppose that, beyond $f^{[0,0]}=1$, now also $f^{[1,1]}=\mathcal{G}_i$, with $\mathcal{G}_i$ a known numerical vector with some non-zero entries.
Once again, this represents an excitation with the same parity in terms of grade, and with an amount of anisotropy as ingredient.
Then, $p^{(0)}$, $q^{(0)}$ and $q^{(1)}$ are the same as in subsection \ref{c1}, with equations (\ref{nought})-(\ref{20})-(\ref{22}) still holding.
However, now (\ref{11}) takes the form:
\begin{eqnarray*}
 \left(\partial_t+\vec{u}\cdot\vec{\nabla}-D\nabla^2+\frac{1}{\tau}\right)p^{[1,1]}_i&=&\left(\frac{Jz_i}{\tau\sigma}-\sigma\nabla_i\right)p^{[0,0]}\\
 &&+\frac{\delta(\vec{x})}{T}\mathcal{G}_i\;.
\end{eqnarray*}
Therefore, even if the expansions for $p^{(1)}$, $p^{(2)}$ and $q^{(2)}$ keep the same form as in subsection \ref{c1}, they are made up by quantities
now obeying different equations. This solves the problem up to $O(\alpha)$.
\subsubsection{Source acting on the opposite parity} \label{c1o}
Let us suppose that, beyond $f^{[0,0]}=1$, now also $f^{[0,1]}=\mathcal{H}_i$, with $\mathcal{H}_i$ a known numerical vector with some non-zero entries.
This represents the addition of an anisotropic excitation with the opposite parity in terms of grade with respect to the basic state,
so that the complete forcing shows a mixed character. Equations (\ref{nought})-(\ref{11})-(\ref{20})-(\ref{22}) still hold, but now one also has
\[\left(\partial_t+\vec{u}\cdot\vec{\nabla}-D\nabla^2+\frac{1}{\tau}\right)p^{[0,1]}_i=\frac{\delta(\vec{x})}{T}\mathcal{H}_i\;,\]
\[(\partial_t+\vec{u}\cdot\vec{\nabla}-D\nabla^2)p^{[1,0]}=-\sigma\nabla_jp^{[0,1]}_j\;,\]
plus three new equations for $p^{[1,2]}_{ij}$, $p^{[2,1]}_i$ and $p^{[2,3]}_{ijk}$ (not reported here).
The problem up to $O(\alpha)$ is solved by noting that $q^{(0)}$ is the same as in subsection \ref{c1}, that the apparent form of $q^{(1)}$ and $q^{(2)}$
is also the same as in there (but the forcing terms for the constituting quantities are different), and that now
$p^{(0)}=(p^{[0,0]}+p^{[0,1]}_iy_i\sigma^{-1})G(\vec{y})$, $p^{(1)}=[p^{[1,0]}+p^{[1,1]}_iy_i\sigma^{-1}+p^{[1,2]}_{ij}H^{\{2\}}_{ij}(\vec{y})]G(\vec{y})$
and $p^{(2)}=[p^{[2,0]}+p^{[2,1]}_iy_i\sigma^{-1}+p^{[2,2]}_{ij}H^{\{2\}}_{ij}(\vec{y})+p^{[2,3]}_{ijk}H^{\{3\}}_{ijk}(\vec{y})]G(\vec{y})$.
As a result, half-integer orders pop up in the physical-space concentration: $q=q^{(0)}+\alpha^{1/2}q^{(1)}+\alpha q^{(2)}+O(\alpha^{3/2})$.
\subsection{Case $h=0$}
If $h=0$ in (\ref{norm}), then $f^{[n,0]}=0$ also for $n=0$. This means that $p^{[0,0]}$ satisfies an unforced advection--diffusion equation
and thus relaxes to a constant, given by a spatial average of (\ref{in}): $q^{\star}\equiv\langle q_*(\vec{x})\rangle_{\vec{x}}$.
Therefore, let us analyze again each of the cases from the previous list in this new optics.
\subsubsection{Source acting on a higher Hermite sector}
Here, with respect to subsection \ref{c1h}, we have $p^{(0)}=[q^{\star}+p^{[0,2]}_{ij}H^{\{2\}}_{ij}(\vec{y})]G(\vec{y})$
after a possible initial transient. Therefore, the non-trivial part of the physical-space concentration is a small quantity proportional to $|1-\beta|$:
namely, $q-q^{\star}=\alpha q^{(2)}+O(\alpha^2)$.
\subsubsection{Source acting on a higher perturbative order}
Here, with respect to subsection \ref{c1p} and neglecting transients, we have $p^{(0)}=q^{\star}G(\vec{y})$.
Therefore, the evolving part of the physical-space concentration is again a small quantity proportional to $|1-\beta|$:
namely, $q-q^{\star}=\alpha q^{(2)}+O(\alpha^2)$.
\subsubsection{Source acting on the opposite parity}
Here, with respect to subsection \ref{c1o}, we have $p^{(0)}=[q^{\star}+p^{[0,1]}_{ij}y_i\sigma^{-1}]G(\vec{y})$.
Therefore, the fluctuating part of physical-space concentration is a small quantity, now however going as $\sqrt{|1-\beta|}$:
namely, $q-q^{\star}=\alpha^{1/2}q^{(1)}+\alpha q^{(2)}+O(\alpha^{3/2})$.
\section{Conclusions and perspectives} \label{concl}
We have analyzed the evolution of the distribution, both in the phase space and in the physical space,
of inertial particles released by a spatially-localized source and advected by an incompressible flow.
The source has been modeled as punctual for a continuity with previous similar studies in the scientific literature
\cite{MAM11,CMAM2006,CMAM2007,CMAM07,T18}, and because this represents the paradigm of inhomogeneity and the basis
for the decomposition of every linear differential problem in terms of Green functions.
Nevertheless, we underline that, for our procedure to hold, such a modelization is not strictly necessary.
Indeed, any source (\ref{source}) of the type $S(\xvt)=F(\xt)f(\vec{v})$ could work, as long as the decomposition
consists in a multiplication where the factor from covelocity is always the same for any space and time.
Since we have been able to deal with situations of both actual release ($h=1$) and coexisting emission/absorption ($h=0$),
we stress that the latter case is particularly significant for instance when a line in $d=2$ or a sheet in $d=3$
represent a real source of particles with outgoing covelocity and a sink for incoming impurities,
at least when the fluid velocity is such as to allow for finding an easy interpretation of the relation between particle velocity and covelocity.

The expansion into a power series in the --- small --- mass-density discrepancy between fluid and particles,
followed by a projection onto the sectors associated with multivariate Hermite polynomials
(where the variance of the corresponding Gaussian measure is dictated by the parameters of the small-scale particle dynamics and equilibrium),
induces a complete separation of the covelocity degree of freedom from the space-time ones.
It is worth mentioning that such an expansion is a regular perturbative one,
free from the multiple-scale character which often faces problems of secular instabilities
and imposes restrictions on the class of tractable flows (periodicity or steadiness, here absent).
We have then derived a set of equations of the advection--diffusion type, possibly with a reaction term,
which are easily solvable at least numerically because living in the physical space,
with a drastic reduction in the dimensionality of the problem.
As a future perspective, it would be interesting to attack this system of equations analytically
for specific instances of flows, such as cellular/(anti)symmetric \cite{MAG17} or parallel \cite{MAMGM17} ones.
Further developments could consist in the inclusion of effects such as flow compressibility
or of the corrections to (\ref{basic}) due to the Fax\'en, Basset, Saffman and Oseen contributions.
\section*{Authors' contributions and acknowledgments}
All the authors were involved in the preparation of the manuscript.
All the authors have read and approved the final manuscript.

%\section{Acknowledgments}
\begin{acknowledgement}
This article is based upon work from COST Action MP1305 ``Flowing Matter'', supported by COST (European Cooperation in Science and Technology).
We thank Andrea Mazzino, Paolo Muratore-Ginanneschi and Semyon Yakubovich for useful discussions and suggestions.
The authors were partially supported by CMUP (UID/MAT/00144/2013), funded by FCT (Portugal) with national (MCTES)
and European structural funds (FEDER), under the partnership agreement PT2020 - ext.\ 2018,
and by Project STRIDE NORTE-01-0145-FEDER-000033, funded by ERDF NORTE 2020.
\end{acknowledgement}

\bibliographystyle{epj}
\bibliography{psdqnbipBIB}

\end{document}